\documentclass[preprint,review,final]{elsarticle}
\usepackage{amsmath}
\usepackage{amsfonts}
\usepackage{lineno,hyperref}
\usepackage[dvipsnames]{xcolor}
\usepackage{subcaption}
\usepackage{xcolor}
\usepackage[utf8x]{inputenc}
\usepackage{chngcntr}
\usepackage{siunitx}
\usepackage{soul}
\counterwithout{figure}{subsection}
\usepackage{changes}
\usepackage{mathtools}

\def\bfO{{\mathbf{y}}}
\def\bfIn{{\mathbf{x}}}
\def\bfR{{\mathbf{R}}}
\journal{Nuclear Inst. and Methods in Physics Research, A}

\makeatletter
\def\ps@pprintTitle{
 \def\@oddfoot{}%
 \let\@evenfoot\@oddfoot}
\makeatother

\begin{document}


\begin{frontmatter}


\title{Quantitative and Three-Dimensional Assessment of Holdup Material}

\author[mymainaddress]{N. Rebei\corref{mycorrespondingauthor}}
\cortext[mycorrespondingauthor]{Corresponding author. Tel.: +1 217 904 6253. Fax.: +1 217 333 2906}
\ead{nrebei2@illinois.edu}

\author[mymainaddress]{M. Fang}

\author[mymainaddress]{A. Di Fulvio}

\address[mymainaddress]{Department of Nuclear, Plasma, and Radiological
                        Engineering, \\University of Illinois, Urbana-Champaign,
                        \\ 104 South Wright Street, Urbana, IL 61801, United
                        States}

\begin{abstract}
    Nuclear material deposited in equipment, transfer lines, and ventilation systems of a processing facility is usually referred to as holdup. Operators need to know the location and amount of holdup in their facility for radiation safety and material accountability. In this work, we propose to use an array of detectors co-axial to the inspected pipe to measure the holdup material. This method is implementable into an automated system capable of crawling on surfaces and pipes of various curvatures, which would enable faster, easier, and more accurate holdup safeguards measurements.
    We first demonstrated that the current holdup assay procedure could lead to a non-negligible bias in the estimate of special nuclear material mass, due to the simplified assumption of deposited geometry introduced by the Generalized Geometry Holdup (GGH) model. The new approach consists of imaging the inner holdup material by characterizing the detector array's response and unfolding it from the measured light output. Our experimental proof of principle consists of three NaI(Tl) detectors surrounding an aluminum pipe containing two cesium-137 (\added{$^{137}$Cs}) sources. We derived the source distribution inside the pipe by first calculating the detector response matrix \added{using a method adaptive to the surface geometry of the object containing the measured holdup material.} Creating a matrix of the detector array's measured counts, we then proceed to solve an inverse problem, resulting in an accurately located source position \added{and activity distribution within the response matrix's spatial resolution}. 
    We then developed a simulated model of the envisioned experimental setup, which accurately described both the activity and position of the source in 2D. Finally, we extended our model onto a discretized three-dimensional model of the system, encompassing 36 detectors. The model was simulated to measure the source distribution and activity of multiple sources in \added{pipes of varied geometries}, which could then be conveniently translated to a real-time holdup measurement protocol. For the 3D simulation of four different source geometries, the model accurately localized the source position in 3D, while the activity retained a maximum relative error of \added{$\pm5.32\%$}. %
\end{abstract}


\end{frontmatter}

\section{Background and Motivation}
    For radiation safety and material accountability, operators need to know the location and amount of hold up, i.e., nuclear material deposited in various systems of a fuel reprocessing facility. Gamma-ray spectroscopy is typically used to estimate holdup activity accurately and requires the use of collimated and shielded gamma-ray detectors, typically inorganic scintillators.
    
    Procedures including quantitative measurements of holdup benefit from assumptions of specific source spatial distribution, enabling operators to perform and analyze thousands of measurements. These procedures are based on the Generalized Geometry Holdup model (GGH)~\cite{russo2005gamma}. However, the bias introduced from the use of the GGH model can lead to a material loss trend spanning years to decades, which may result in adverse regulatory and financial issues~\cite{resolution}.
    
    The manual measurement of holdup in piping is a costly, time consuming, and hard-working process. For this reason, automated robotic systems have been proposed and tested~\cite{jones2019robot}. However, the existing systems are destructive, requiring the operator's physical intervention to insert the robot inside the pipe manually. These current systems are optimized for a single diameter and are not designed to inspect pipes of various curvatures. Furthermore, other types of equipment, such as filters, require extensive modification to be inspected by such robots.
    
    We propose a soft robot that is minimally tethered~\cite{softrobot2} and is equipped with radiation detectors and ultrasound sensors, capable of crawling on surfaces and adjusting to the curvature of the pipe's geometry without the need of \added{re-positioning the robot. The operator could use a telescopic rod to initially place the system at the desired location and then remotely control its movement or let the device autonomously reach the position and perform the measurements. This robot would allow an accurate localization and non-destructive characterization of the holdup material, making measurements faster, more accurate, autonomous, and less demanding on operators.}

    In this paper, we focus on the imaging reconstruction and material quantification algorithm, which will result in a 3D mapping and activity of the deposition profile. We will first discuss and present the bias introduced in the GGH model, then propose a new approach to analyze data that can be acquired by such a system. The proposed approach consists of characterizing the detector array's response and solving a linear inverse problem to derive the activity of the holdup. This approach's accuracy is tested in an experimental setup, then in simulations performed using MCNP6.2~\cite{osti_1419730} in 2D and 3D cases.
    
\section{Methods}
\subsection{GGH Approach}
Holdup analysis usually involves measuring spectra that have overlapping peaks in energy regions between 200 and 800 keV. Thus, the calibration and measurement efforts must be simple, and the analysis must apply to the unique geometry of each measurement. This objective is achieved by adjusting holdup measurement geometries in the field of view of a cylindrical collimated gamma-ray detector so that each holdup deposit geometry can be generalized as a point, line, or area~\cite{russo2005gamma}. After detector calibration, the procedure of holdup measurement follows three steps:
\begin{enumerate}
\item A coarse survey is performed to determine if the deposited geometry can be generalized as a point, line, or area. 
\item An accurate measurement of the deposited material is acquired.
\item An analytical model is applied to interpret the data according to the chosen holdup geometry in step 1.
\end{enumerate}

\begin{figure}[ht]
    \centering
    \includegraphics[width=0.75\linewidth]{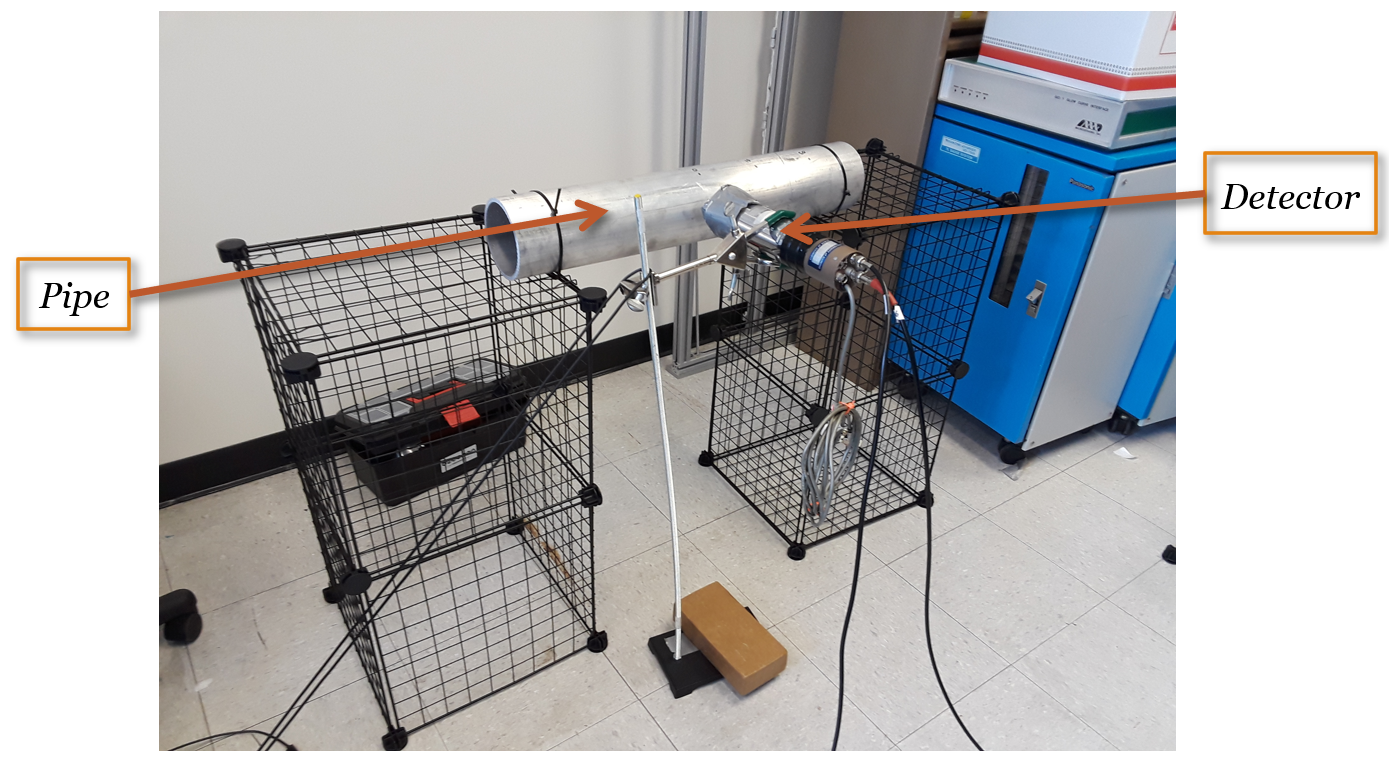}
    \caption{Setup of single detector measurement with pipe and detector labeled.
}
    \label{fig:GGH measurement}
\end{figure}

We first reproduced a holdup measurement using laboratory sources to identify the maximum bias due to a misidentification of the source's geometry. We used two 1~$\mu$Ci \added{$^{137}$Cs} point sources at the same location and separated the two sources to investigate the misidentification of the source activity due to the distance. \added{We followed the standard holdup measurement procedure, where a coarse survey of the equipment of interest is performed prior to holdup assessment.} Using a survey process based on measurements from a Victoreen 450 survey meter, we determined that when the sources were located within up to 10~cm from each other, they were identified as a single source.
Thus, the two sources were placed 0, 2, 4, 6, 8, and  10~cm apart. \added{At the expense of a longer measurement time, the use of a collimated, high-efficiency detector instead of the Victoreen survey meter would likely allow to discriminate the two sources even when closer than 10 cm to each other.} We then set a NaI(Tl) detector next to a pipe containing the two \added{$^{137}$Cs} sources to simulate the current method of an operator holding up a detector towards the source, as shown in Fig.~\ref{fig:GGH measurement}. \added{The pipe to detector center distance was chosen to resemble a realistic method of holdup assay where it is not possible to optimize the pipe detector distance based on the source shape estimate.}

We measured the distance between the center of the detector and the assumed point source location and the counts at each distance. Based on the initial survey, the sources would be identified as a point source. Thus, the activity $A_H$ of the \added{two $^{137}$Cs sources} was then calculated for each measurement using a point model as follows:
\begin{equation}
    A_H = K_pC_Hr^2_H,
\end{equation}
where $r_H$ is the measurement distance between the center of the detector and the point source, $C_{H}$ is the corrected net count rate of the gamma peak, $A_{H}$ is the activity of the point source, and $K_p$ is the calibration constant.

\begin{figure}[h!]
\centering
\includegraphics[width=1\linewidth]{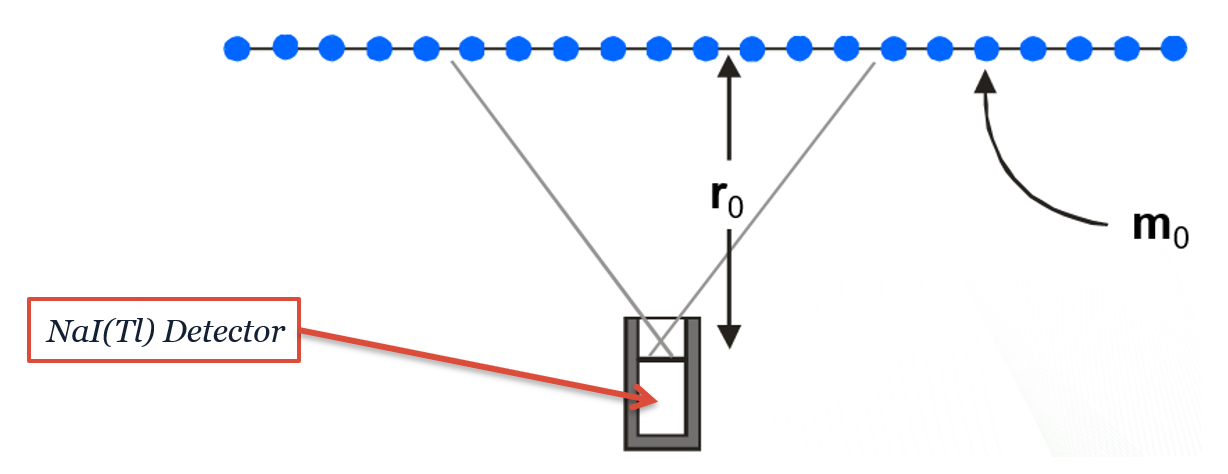}
\caption{\added{Detector responses for the source with mass $m_0$ are measured at successive points along a line perpendicular to the detector axis at a known distance $r_0$ from the detector face.}}
\label{fig:calibration}
\end{figure}

To obtain the calibration constant $K_p$, we measured a \added{$^{137}$Cs} point source of known activity at multiple positions along a line \added{perpendicular to the detector axis} following the standard procedure~\cite{russo2005gamma}, \added{as seen in Fig.~\ref{fig:calibration}. More precisely, the standard procedure is designed to allow the operator to obtain a detector response scalable from one dimension to two dimensions by moving the source to detector distance in a linear pattern. Further details on the standard approach can be found in reference~\cite{russo2005gamma} at \textbf{VIII.1.}}

For each measurement, the calibration constant $\overline{K_p}$ can be calculated as
\begin{equation}
    \overline{K_p} = \frac{A_{cal}}{C_{cal}r^2_{cal}},
\end{equation}
where $C_{cal}$ is the corrected net count rate of the gamma peak, $A_{cal}$ is the known activity of the point source, and $r_{cal}$ is the distance between the center of the detector and the source. \added{The average of the calculated calibration constants $\overline{K_p}$ is then determined to be used as $K_p$ in Eq. 1 for the single detector measurement.}


\subsection{Linear Inverse Approach}
Suppose the holdup deposited inside a cross section of a pipe is contained in a arbitrary \added{bounding box} ${\Omega}$. Its distribution can be reconstructed by solving an inverse problem where the detector response and the measured counts are known, the solution yielding the most likely source distribution to produce the measured counts. 

We first discretize the region of interest ${\Omega}$ into $N$ 1~cm side square pixels and we would like to know the distribution of the radioactivity expressed as $\mathbf{\bfIn}$ in $\Omega$ as a function of the pixel number. We measure the counts $\mathbf{y}$ using a detector array encompassing $M$ detectors and assume the system to be linear:
\begin{equation}
    \mathbf{y}  = \mathbf{R} \mathbf{x} ,  \mathbf{y} \in \mathbb{R}^{M\times1},\mathbf{x} \in \mathbb{R}^{N\times1}, \mathbf{R} \in \mathbb{R}^{M\times N},\label{eq:linear_eq}
\end{equation}
where $\bfR$ is the system response matrix. $\bfR_{i,j}$ are the counts of detector $i$ when there is a source of unit activity in pixel $j$, which could be obtained by direct calculation or simulation.

For direct calculation, the response matrix $\bfR$ is calculated based on the gamma-ray attenuation law:
\begin{equation} \label{eq:3}
 \bfR_{i,j} = f \omega_{\text{int}} \omega_{\text{geo}} e^{-\mu_a d_{a}} e^{-\mu_w d_{w}}
\end{equation}
$f$ is the branching ratio of the 662~keV \added{$^{137}$Cs}, \textit{i.e.}, 0.85. $\omega_{\text{int}}, \omega_{\text{geo}} $ is the intrinsic efficiency and geometric efficiency of detector $i$, respectively. $\mu_a,\mu_w$ are the attenuation coefficients of the 662~keV gamma rays in air and pipe wall, respectively. $d_a,d_w$ are the distances that the gamma rays travelled in air and pipe wall, respectively. If the pixel center is outside the pipe's cross section, the responses from each detector will be assumed to be $0$. 

\added{Regardless of the measured pipe cross-section's geometry, the described procedure of creating the response matrix will not change. It would suffice to enter the thickness of the pipe or any other changing parameter to the model to obtain a new response. If the source-to-detector distances or number of pixels are changed due to a different pipe diameter or altered detector positions, the response matrix will also need to be recalculated. Given the dimensions of the pipe are well mapped out, the response matrix will fit to the geometry of the pipe's cross section within its spatial resolution. The spatial resolution of the response can be adjusted by changing the dimensions of the pixels.}

Given a $M\times N$ design matrix $\bfR$ and a target matrix $\bfO$ with $M$ elements, we can solve Eq.~\ref{eq:linear_eq} for the source distribution $\mathbf{x}$. However, this linear system is usually under-determined because $M << N$. We solve the following optimization problem to get the minimum mean square error estimate of $\mathbf{x}$:
\begin{equation}
    \hat{\bfIn} = \arg \underset{\bfIn}{\min} \{\|\bfR \bfIn - \bfO\|^2 + \lambda||x||_1\}, \bfIn_i \geq 0, i  \in  \mathbb{Z}  \cap  [1,N],\label{eq:mmse}
\end{equation}
 where $||x||_1$ stands for the sum of the absolute values of the components of $\bfIn$. We use a Fast Iterative Shrinkage-Thresholding Algorithm (FISTA) to perform the minimization~\cite{fista2009}, with $\lambda = 10^{-7}$. The minimization process terminates when the error $\|\bfR \bfIn\ - \bfO\|$ converges, i.e., two subsequent errors have a relative change less than $10^{-4}$. \added{FISTA was chosen due to its favorable computational time for an operator in a facility setting.}



\subsection{Experimental Validation of a 2D case in a Simplified Geometry}

\begin{figure}[h!]
\centering
\includegraphics[width=1\linewidth]{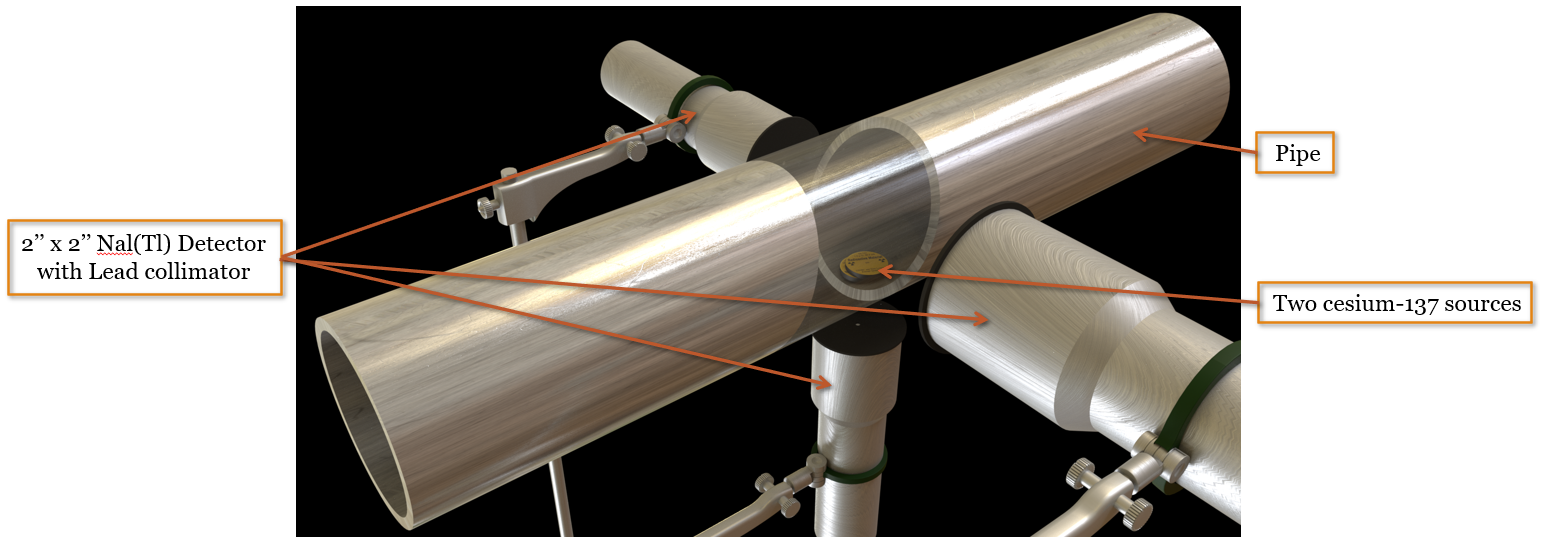}
\caption{3D sketch of the experiment. Center of pipe made transparent to show two \added{$^{137}$Cs} sources.}
\label{fig:exp_sketch}
\end{figure}

Fig.~\ref{fig:exp_sketch} shows the experimental setup that we used to validate the linear-inverse approach. Two 
\added{$^{137}$Cs} sources with a total activity of 70.80 $\pm$ 14.16~kBq~\added{1SD}  are stacked on one another at the center of an aluminum cylindrical pipe of thickness 0.6~cm.
Three 2''~$\times$~2'' NaI(Tl) detectors are then positioned around the center of the pipe. Lead collimators with an aperture of 8~mm and thickness 1.79~cm are used for each detector, which are able to attenuate up to 90\% of the gamma rays.

We measured the gamma-ray spectra for 60 minutes, then calculated the net counts of the 662-keV peak by subtracting the background and Compton continuum from the raw spectra. The collection of the net counts from the three detectors is the target matrix $\bfO$.

\begin{figure}[ht]
    \centering
    \includegraphics[width=0.5\linewidth]{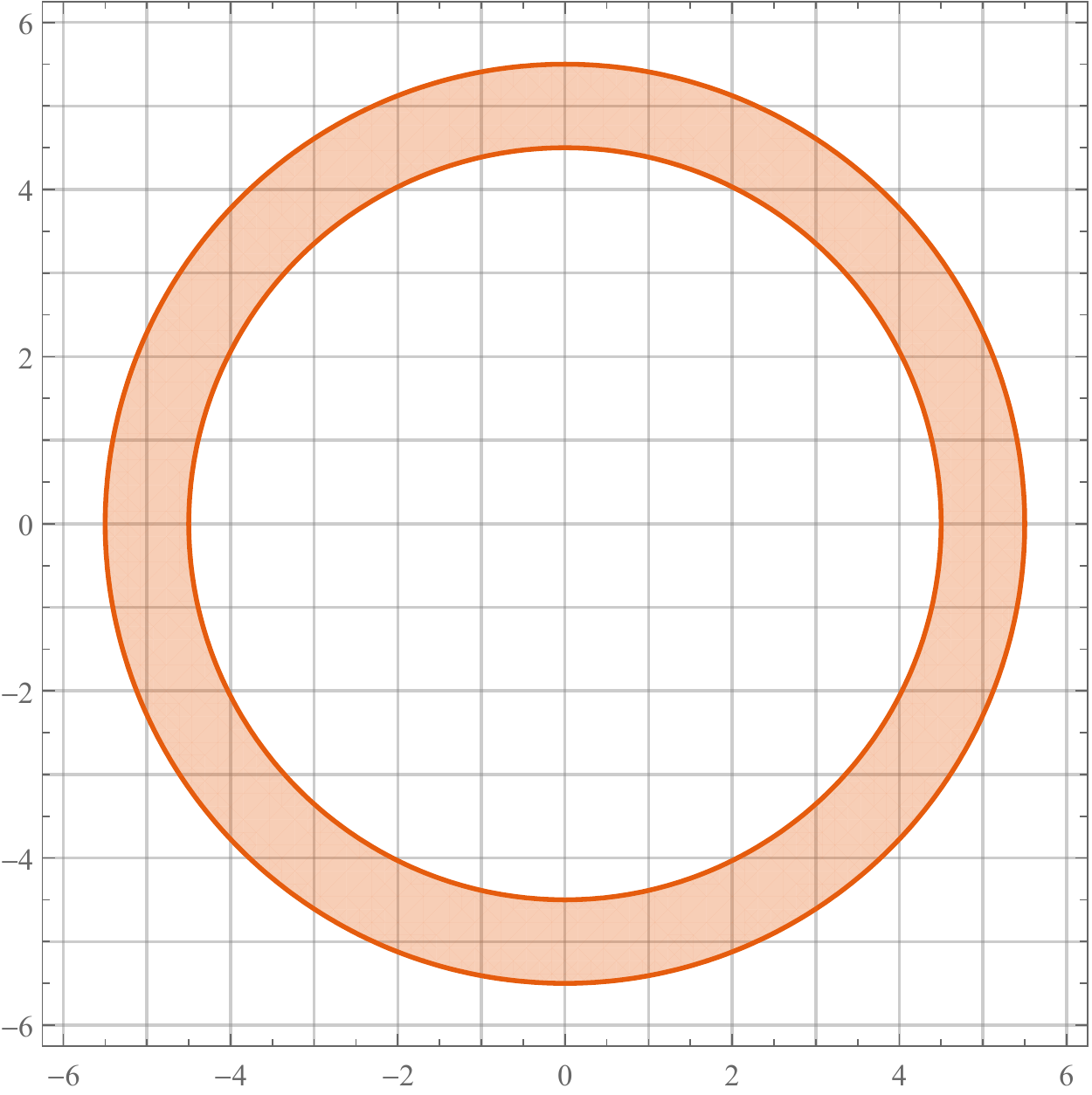}
    \caption{2D discretization of the pipe cross section.}
    \label{fig:2d_pipe}
\end{figure}

As the outside diameter of the pipe is 11.4 cm, the region of interest in this experiment is a square of area $12\times12$~cm${}^2$ encompassing the cross section of the pipe. The square is divided into 144 square pixels of area $1\times1$~cm${}^2$, as shown in Fig.~\ref{fig:2d_pipe}. We are interested in the source activity $\bfIn_i$ in each pixel, $i=1,\cdots, 144$.

\added{The geometric efficiency is determined for each detector based on the source-detector geometry in the experimental validation, given by the fractional solid angle subtended by the detector crystal at the assumed point source: }
\begin{equation}
    \omega_{\text{geo}} = \frac{1}{2}\left(1- \frac{d}{\sqrt{d^2 + r^2}} \right),
\end{equation}
\added{where $d$ is the source-detector distance and $r$ is the detector radius.
Since the NaI(Tl) detector is a cylinder, the distance $d$ to the source is not uniform. We use the average distance $\overline{d}$ to the detector instead, given by}
\begin{equation}
    \overline{d} = \sqrt{\frac{r^2}{2} + \frac{h^2}{12} + c^2},
\end{equation}
\added{where $h$ is the detector length, $r$ is the detector radius and $c$ is the distance from the source to detector center~\cite{knoll2010radiation}.}


\added{The intrisnic efficiency is then determined for each detector using a single measurement with a \added{$^{137}$Cs} source at a known position from the detector. The intrinsic peak efficiency is given according to the standard procedure in reference~\cite{knoll2010radiation} by}
\begin{equation}
    \omega_{\text{int}} =  \frac{N}{\omega_{\text{geo}} \times BF \times BR \times T \times A},
\end{equation}
\added{where $N$ is the measured counts in the peak area, $\omega_{\text{geo}}$ is the fractional solid angle between the measured source and detector, $BF$ is the branching fraction, i.e., 100\%, $BR$ is the branching ratio, i.e., 85.10\%, $T$ is the acquisition time of the detector, and $A$ is the activity of the measured source in Bq.}

\added{In our case, the calibration source and the experimental test source are both $^{137}$Cs, with the \added{$^{137}$Cs} source used as one of the two sources in the experimental validation measurement. Thus, the intrinsic efficiency factored into the response matrix matches the actual measurement intrinsic efficiency. This favorable condition may not be possible in a real world case where a weighed efficiency as a function of the energy would be used in the response matrix. }

The system response matrix $\bfR$ is then calculated using Eq.~\ref{eq:3}. 
We solve the inverse problem for $\mathbf{\bfIn}$ using Eq.~\ref{eq:mmse} of our inverse linear approach with $M = 3$ and $N = 144$, i.e., $\bfR \in \mathbb{R}^{3\times 144}$, $\bfO \in \mathbb{R}^{3 \times 1}$, and $\bfIn \in \mathbb{R}^{144 \times 1}$. The calculated activity in kBq of the source inside the pipe is then given by
\begin{equation}
\begin{aligned}
    A_{cal} = \frac{\sum_{i=1}^{N} \bfIn_i}{t} \times \frac{1 \text{ kBq}}{1000 \text{ Bq}},
\end{aligned}
\end{equation}
where $t$ is the measurement time in seconds. 

\subsection{Validation via Simulation}
We simulated a two-dimensional (2D) and three-dimensional configuration to examine the inverse approach in greater detail. The simulation is performed using MCNP6.2~\cite{osti_1419730}. We simulated both $\bfR$ and $\bfO$ by recording the detector response and output to a source placed in the pipe. $\mathbf{\bfIn}$ is finally calculated using our linear inverse approach. \added{The statistical counting error is not propagated in the reconstruction.}

\subsubsection{2D simulation}
We simulated the experiment with the same geometry and detector positions in our experimental validation, using a single \added{$^{137}$Cs} source with a source strength of $10^8$ gamma rays per second. We similarly obtained the response matrix containing 144 square pixels by simulating the detector response to a source of unit activity in each pixel. Given the simulated counts $\bfO \in \mathbb{R}^{3 \times 1}$ and responses $\bfR \in \mathbb{R}^{3\times 144}$ from the three detectors, our goal is to determine the activity in each pixel $\bfIn \in \mathbb{R}^{144}$ and reconstruct a 2D image of the source distribution by solving for the desired source distribution using the linear inverse approach.
    
\subsubsection{3D simulation}
\begin{figure}[!htbp]
\captionsetup{font=footnotesize}
    \centering
    \begin{subfigure}{0.8\linewidth}
    \captionsetup{font=footnotesize}
        \centering
        \includegraphics[width=0.65\linewidth]{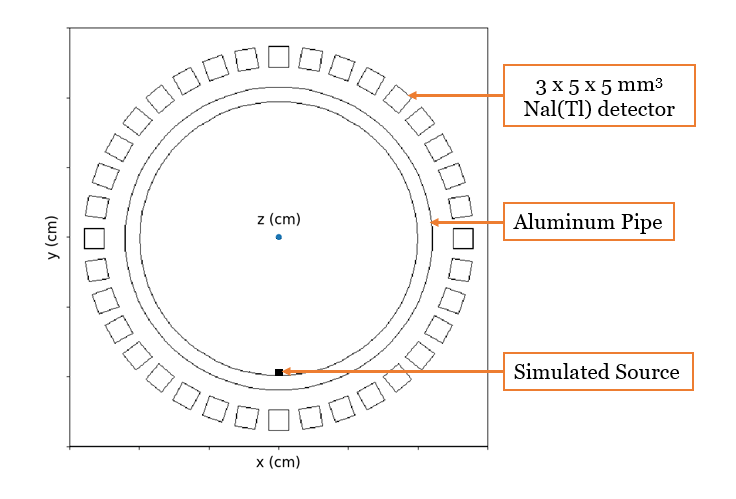}
        \caption{Source 1} 
        \label{fig:3d_simu_detector1}
    \end{subfigure}%
    \newline
    \begin{subfigure}{0.9\linewidth}
    \captionsetup{font=footnotesize}
        \centering
        \includegraphics[width=0.7\linewidth]{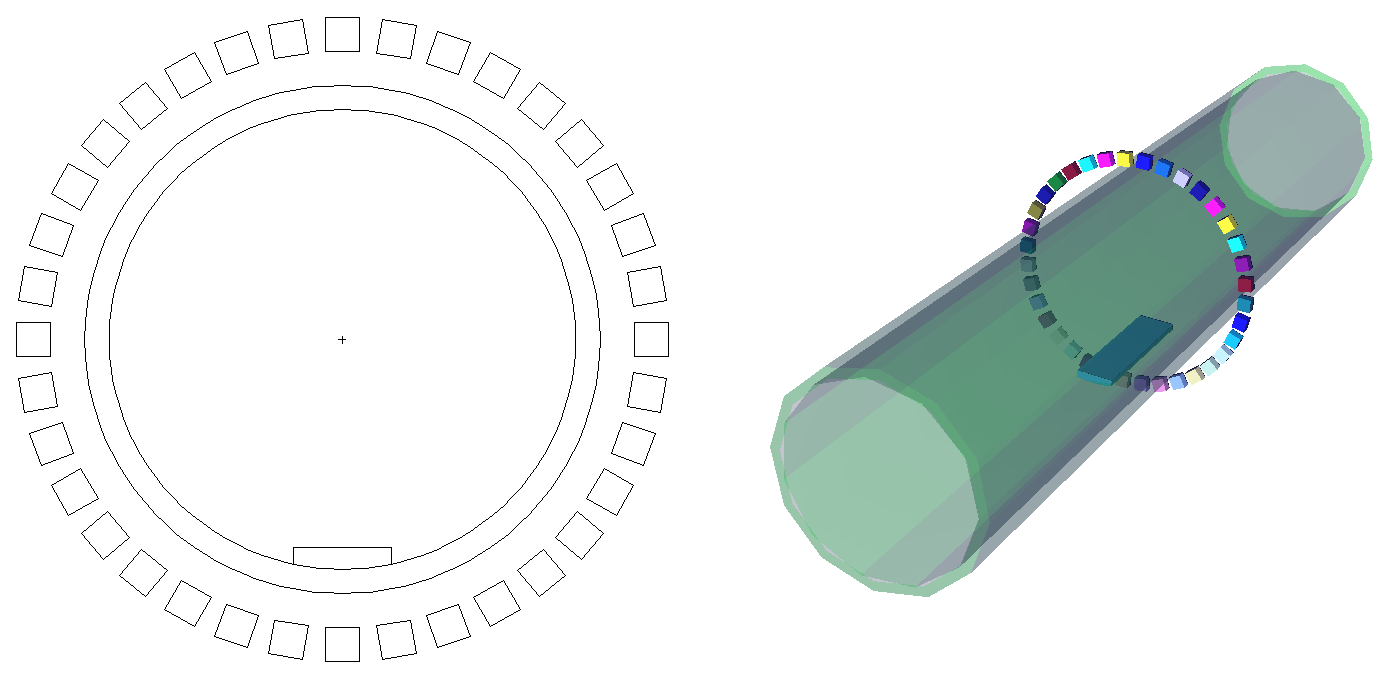}
        \caption{Source 2}
        \label{fig:3d_simu_detector2}
    \end{subfigure}
    \newline
    \begin{subfigure}{0.9\linewidth}
    \captionsetup{font=footnotesize}
        \centering
        \includegraphics[width=0.7\linewidth]{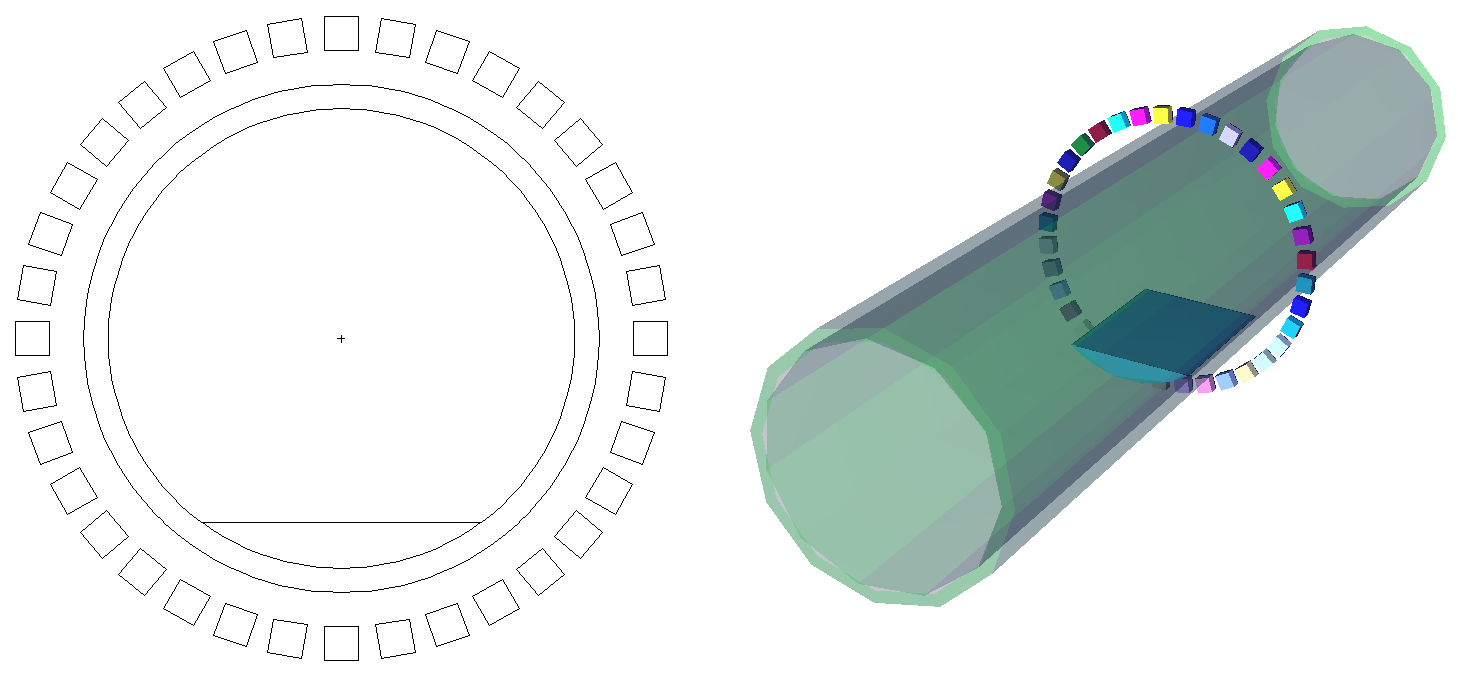}
        \caption{Source 3}
        \label{fig:3d_simu_detector3}
    \end{subfigure}%
    \newline
    \begin{subfigure}{.5\linewidth}
    \captionsetup{font=footnotesize}
        \centering
        \includegraphics[width=1\linewidth]{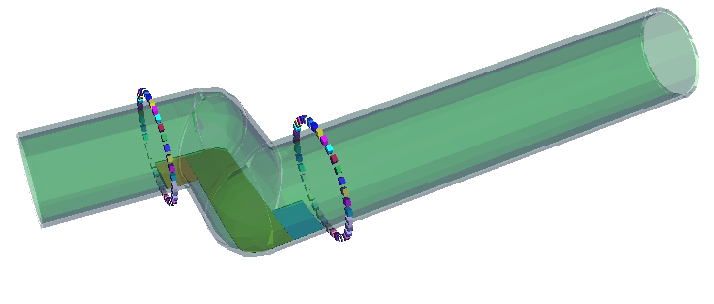}
        \caption{\added{Source 4. Detector rings indicate the starting and ending position of the measurements.}}
        \label{fig:3d_simu_detector4}
    \end{subfigure}%

    \caption{(a,b,c) Cross section of the pipe with source and (b,c\added{,d}) 3D source distribution with source, pipe, and collimated detector ring shown. The detector collimators are not shown. \added{The cross section of the aluminum pipe with source 4 is equivalent to source 3.}}
    \label{fig: 3d_simu}
\end{figure}

In this case, we are interested in the 3D source distribution inside the pipe. We simulated multiple source distributions \added{of uniform density} with an overall source strength of $10^8$ gamma rays per second, with all sources located at the bottom of the pipe:
\begin{description}
\item [Source 1]: A point source located at $z = 0~\text{cm}$, as shown in Fig.~\ref{fig:3d_simu_detector1}.
\item [Source 2]:  A thin wire of square cross section $1\text{cm}^2$ spanning $-4~\text{cm} \leq z \leq 4~\text{cm}$, as shown in Fig.~\ref{fig:3d_simu_detector2}.
\item [Source 3]:  A deposit spanning $-4~\text{cm} \leq z \leq 4~\text{cm}$ of height 1~cm, as shown in Fig.~\ref{fig:3d_simu_detector3}. 
\item [Source 4]: \added{In addition to a standard horizontal pipe, we simulated a deposit in an S-shaped pipe as shown in Fig.~\ref{fig:3d_simu_detector4}.}
\end{description}


    
    

A detector ring containing 36 lead-collimated NaI(Tl) detectors surrounded the pipe and measured the counts at different cross-sections. For sources 1, 2, and 3, the detector ring began at $z = -5$~cm and was moved up to $z = 5$~cm, with the counts recorded in $1$~cm increments. 
\added{In source 4, the detector ring traveled in $1$~cm increments along the horizontal and vertical sections of the pipe, measuring once at each corner with an angle of \ang{45} with respect to the horizontal and vertical. To resemble $1$~cm measurements along the corner, we artificially created eight new measurements at each corner. Four measurements were created between the known corner measurement and the nearest vertical/horizontal measurement, with the estimated measured counts of the detectors as the arithmetic means between the known recorded counts.}

\begin{figure}[!htbp]
    \centering
    \includegraphics[width=0.5\linewidth]{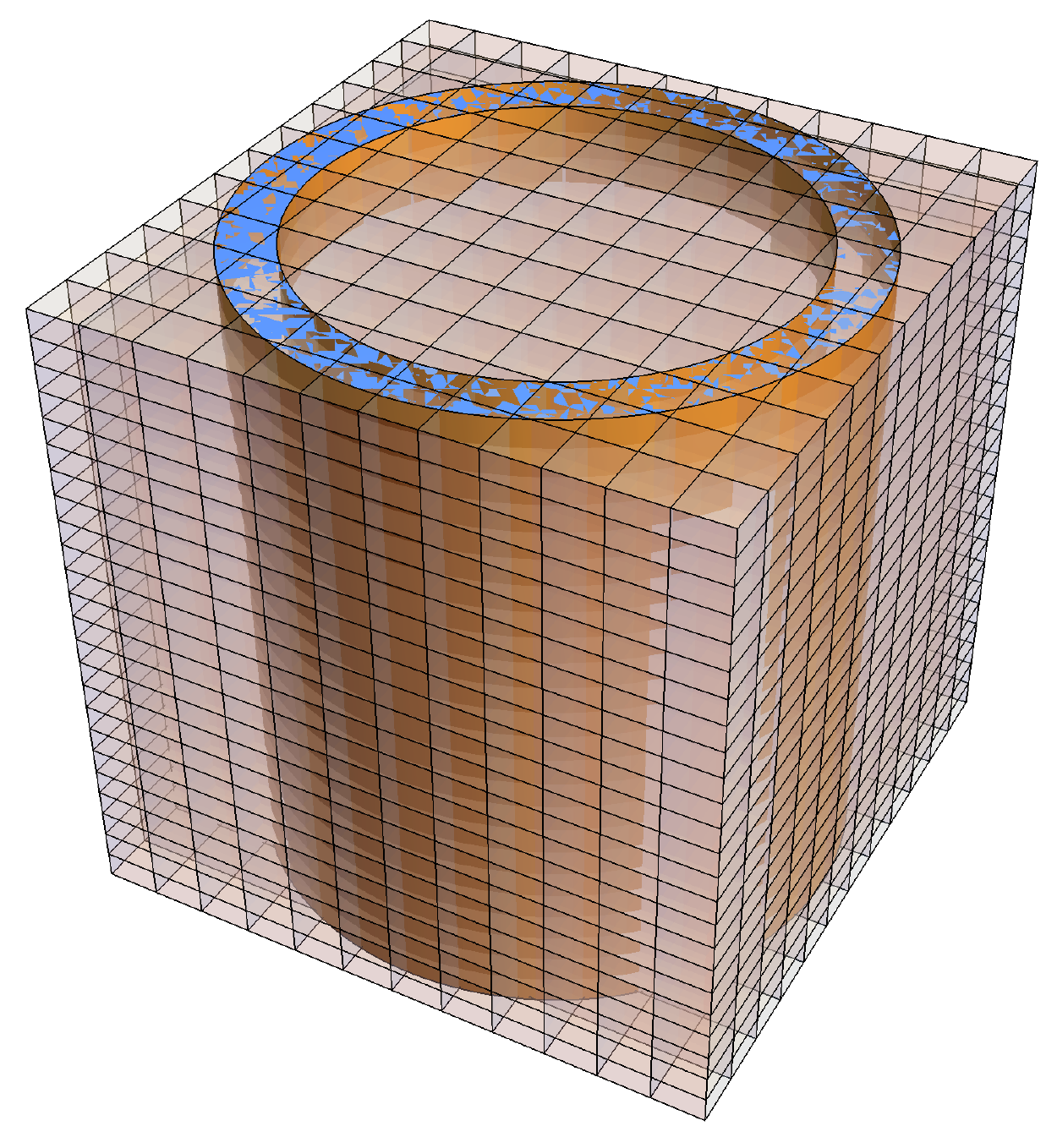}
    \caption{3D discretization of the pipe.}
    \label{fig:3d_pipe}
\end{figure}

To obtain a spatial resolution of the 3D response matrix, we discretized the entire pipe into voxels of 1~cm $\times$ 1~cm $\times$ 1~cm volume, as shown in Fig.~\ref{fig:3d_pipe} for a standard horizontal pipe.

\noindent Let the voxel whose center is 
\begin{equation}
    \begin{aligned}
        &(x_i,y_j,z_k) = (i-6.5, j-6.5,k), \\
        &1\leq i \leq 12, 1\leq j \leq 12, (i,j,k) \in \mathbb{Z}^3
    \end{aligned}
\end{equation}
be labeled by $(i,j,k)$ and $\mathbf{M}_n(i,j,k)$ be the response of the $n$th detector at $(x_n, y_n,0)$ to a source of unit activity at voxel $(i,j,k)$. In the creation of the response matrix, we assume that the detector response only depends on the relative distance between the voxel center and the detector front face. Thus, the response of detector $n$ at $(x_n,y_n,l)$ for the voxel $(i,j,k)$ is equivalent to $\mathbf{M}_n(i,j,|k-l|)$. For each detector, a total of $144 \times 10$ simulations are performed to determine the detector's response at each cross-section of the pipe up to 9~cm away. Therefore, we further assume the response for $|k-l| > 9$ is effectively 0 due to the long distance between the source and detector. We then proceed to solve for the source distribution in the measured cross-sections of the pipe using the linear inverse approach.
\added{This method of discretization was assumed to be true for all pipes simulated, regardless of actual geometry.}

\section{Results}
\subsection{Bias Introduced in the GGH model}
According to the standard procedure, we followed the preliminary measurement which resulted in the incorrect 
We calculated the source mass based on the point model for each distance between two \added{$^{137}$Cs} sources and compared it to the actual mass, as shown in Table~\ref{table:mass estimate point model}. The assumption that the source deposition geometry is a point leads to an underestimate of source mass up to 30\%, and this effect can be seen to increase as the source-to-source distance increases. \added{This underestimate may additionally be partially due to the source-to-detector distance of approximately 10 cm was not optimized to perform a measurement in good geometry~\cite{knoll2010radiation}.}
\begin{table}[!htbp]
    \caption{Percent Differences between Calculated Activity and Real Activity.}
    \label{table:mass estimate point model}
    \centering
    \resizebox{1\linewidth}{!}{
        \begin{tabular}{cccc}
            \hline
            \hline
             Source to Source Distance (cm) & \added{Real Activity} (kBq) & \added{Calculated Activity} (kBq) & Percent Difference \\
            \hline
            0 & 70.80 & 68.24 & 3.69\%\\
            2 & 70.80 & 66.22 & 6.69\% \\
            4 & 70.80 & 62.89 & 11.83\% \\
            6 & 70.80 & 60.57 & 15.58\% \\
            8 & 70.80 & 56.31 & 22.80\% \\
            10 & 70.80 & 52.31 & 30.04\% \\
            \hline
            \hline
        \end{tabular}}
\end{table}

\begin{figure}[!htbp]
\centering
    \begin{subfigure}{.5\linewidth}
    \captionsetup{font=footnotesize}
        \centering
        \includegraphics[width=\linewidth]{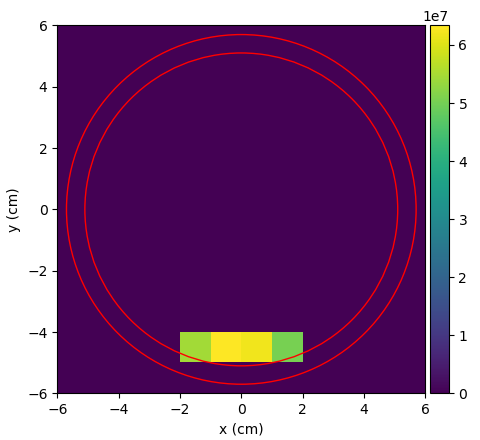}
        \caption{\added{Experiment}}
        \label{fig:2d_result}
    \end{subfigure}%
    \begin{subfigure}{.5\linewidth}
    \captionsetup{font=footnotesize}
        \centering
        \includegraphics[width=\linewidth]{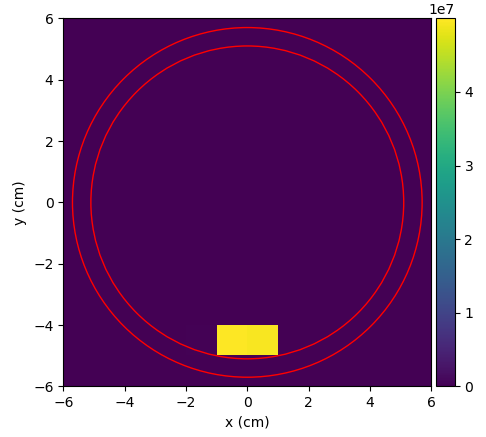}
        \caption{2D simulation}
        \label{fig:2dsim_result}
    \end{subfigure}%
\caption{2D source distribution inside the pipe, i.e. the elements of the calculated $\mathbf{\bfIn}$ on a 2D regular raster. The red circles show the pipe wall cross section. \added{The reconstructed experimental activity distribution exhibits a percentage difference between the green and yellow pixels of approximately 20\%}}
\label{fig:2d_recon}
\end{figure}
\subsection{Experimental Validation in a Simplified Geometry}
Fig.~\ref{fig:2d_result} shows the reconstructed 2D source distribution using FISTA, with the activity estimated as \added{64.00}~kBq.
Compared with the actual source distribution, the source location was accurately determined. Furthermore, the source activity was well within the activity uncertainty provided by the manufacturer, i.e., 70.80±14.16 kBq.


\subsection{Validation via Simulation}
\subsubsection{2D Reconstruction and Activity Quantification}

Fig.~\ref{fig:2dsim_result} shows the reconstructed 2D source distribution based on the simulated detector counts. The activity calculated using FISTA was $99956.15$~kBq, with a relative error of 0.044\%.
Both the source location and activity were accurately determined.

\added{One may notice that the source distribution in Fig.~\ref{fig:2d_result} is slightly wider than Fig.~\ref{fig:2dsim_result}. This artifact may be due to a combination of effects including a slightly incorrect detector and source position assessment, in addition to the choice of iteration parameters in FISTA. As FISTA undergoes more iterations after the stopped iteration count of the experimental validation, the calculated source distribution from Fig.~\ref{fig:2d_result} will become to be only seen in the bottom center two pixels, similar to Fig.~\ref{fig:2dsim_result}.}




\subsubsection{3D Reconstruction and Source Activity Quantification}
\begin{figure}[!htbp]
\captionsetup{font=footnotesize}
    \centering
    \begin{subfigure}{.5\linewidth}
    \captionsetup{font=footnotesize}
        \centering
        \includegraphics[width=1\linewidth]{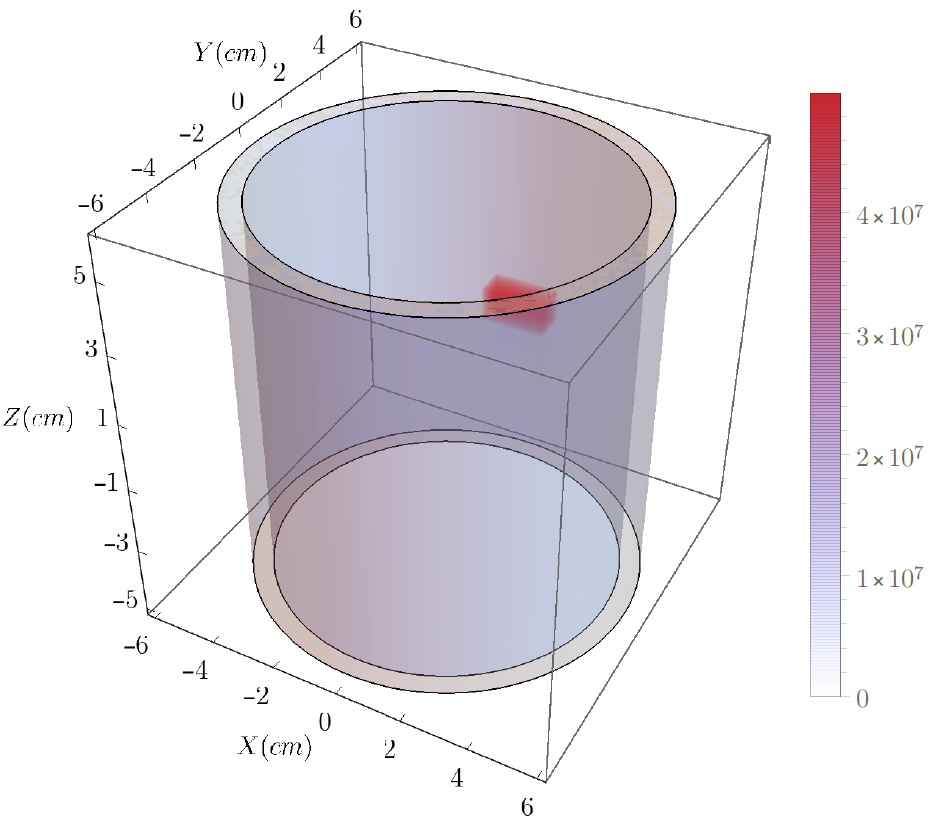}
        \caption{Source 1.}
        \label{fig:3d_resultpoint}
    \end{subfigure}%
    \begin{subfigure}{.5\linewidth}
    \captionsetup{font=footnotesize}
        \centering
        \includegraphics[width=1\linewidth]{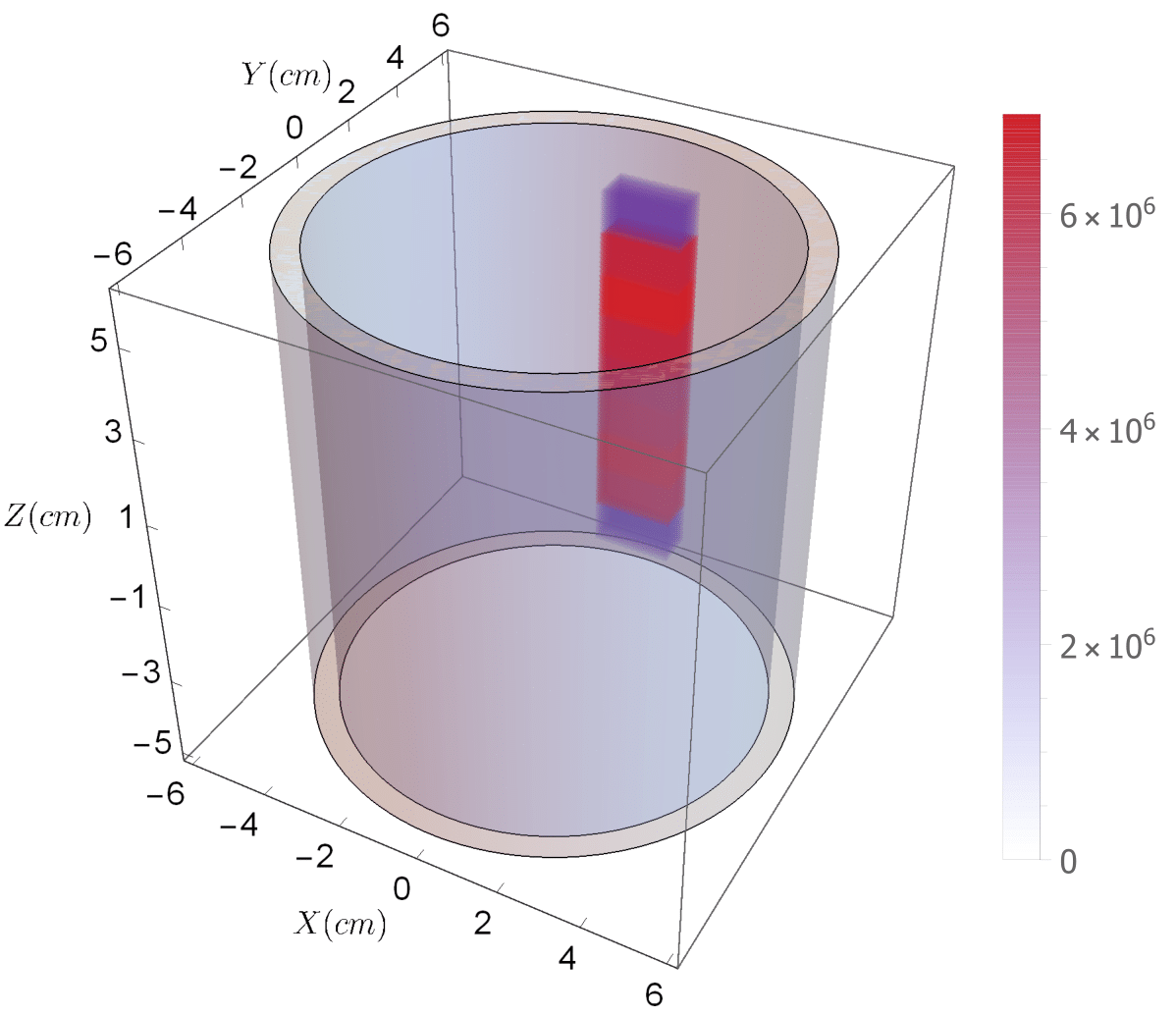}
        \caption{Source 2}
        \label{fig:3d_resultsource2}
    \end{subfigure}
    \newline
    \begin{subfigure}{0.5\linewidth}
    \captionsetup{font=footnotesize}
        \centering
        \includegraphics[width=1\linewidth]{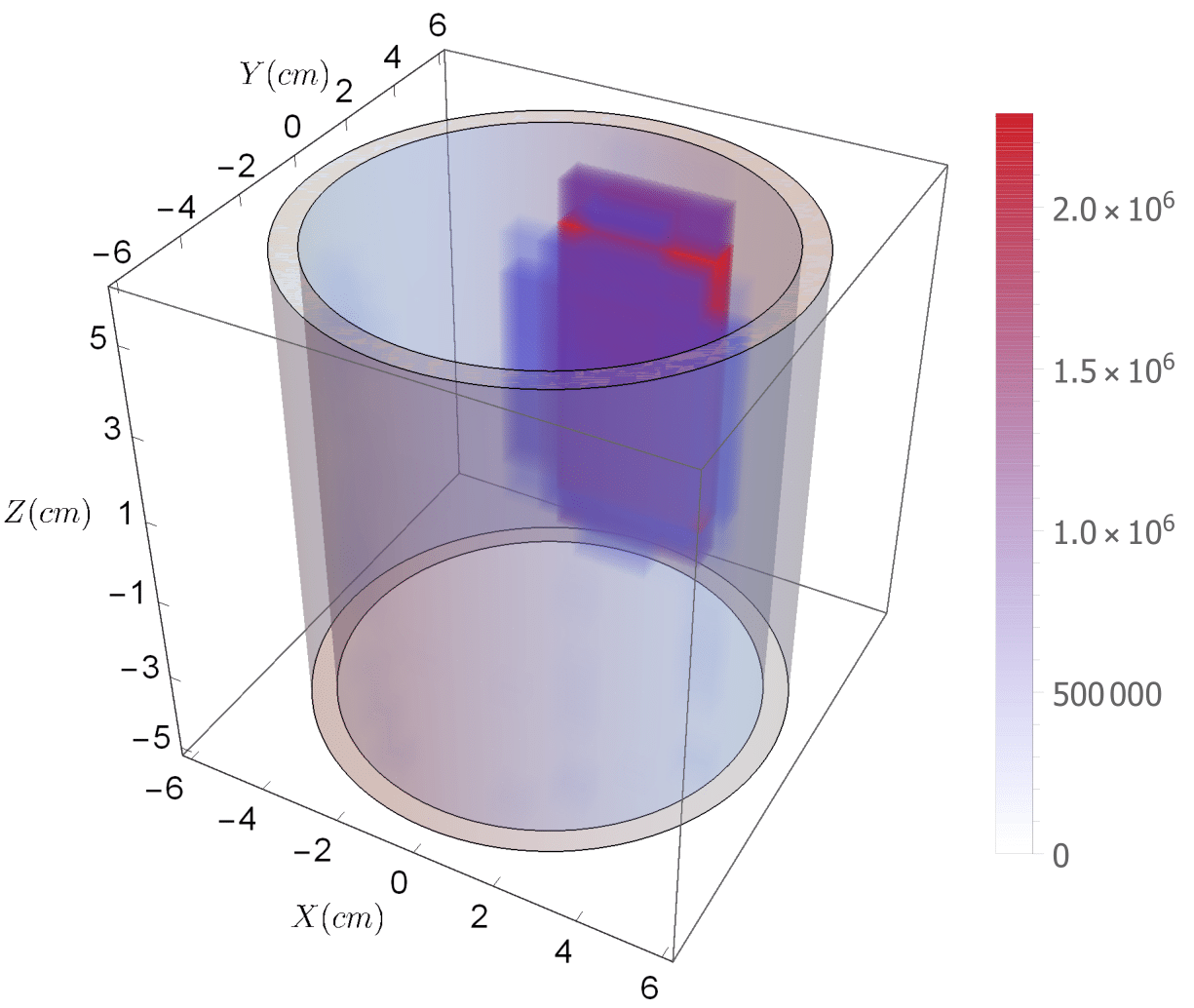}
        \caption{{Source 3}}
        \label{fig:3d_resultsource3}
    \end{subfigure}%
    \begin{subfigure}{0.5\linewidth}
    \captionsetup{font=footnotesize}
        \centering
        \includegraphics[width=1\linewidth]{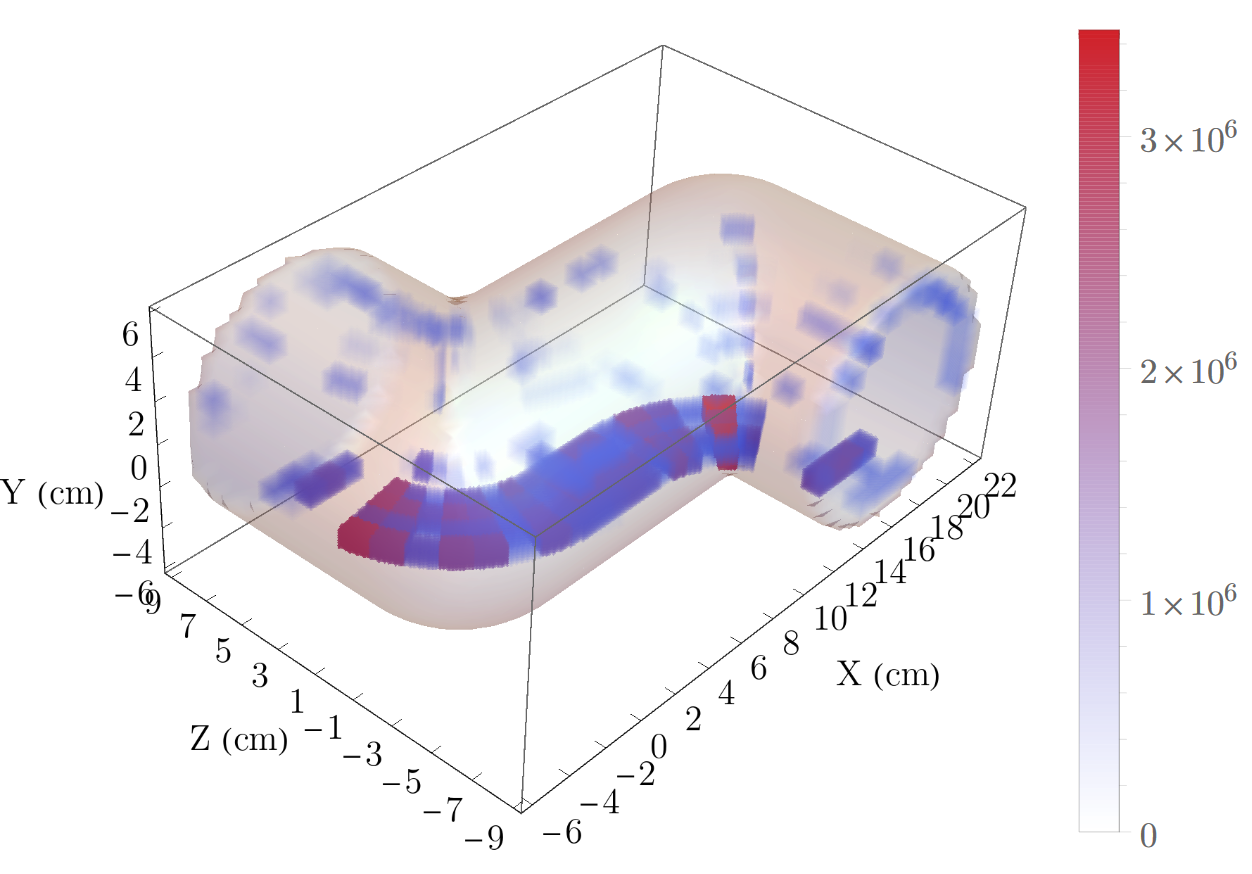}
        \caption{\added{Source 4}}
        \label{fig:3d_resultsource4}
    \end{subfigure}%
    \caption{Reconstructed 3D source distributions.}
    \label{fig: 3d_simu_result}
\end{figure}

We reconstructed the 3D source distributions using FISTA for all four simulated cases, as shown in Fig.~\ref{fig: 3d_simu_result}. The estimated source activities and systematic errors are summarized in Table~\ref{table:activity estimate 3D}. \added{Other than source 4, both the source locations and activities were determined accurately in the 3D reconstructions.}

\begin{table}[!htbp]
    \caption{Comparison between the estimated total activity and actual activity.}
    \label{table:activity estimate 3D}
    \centering
    \resizebox{.7\linewidth}{!}{
        \begin{tabular}{cccc}
            \hline
            \hline
             Source  & Actual activity (kBq) & Estimated activity (kBq) & Relative error \\
            \hline
             1 & 100000 & 101124.78 & 1.12\%\\
            2 & 100000 & 99625.42 & -0.37\% \\
            3 & 100000 & 96081.92 & -3.92\% \\
            \added{4 (FISTA $\lambda = 10^{-7}$)} & 100000 & 105317.99 & 5.32\% \\
            \added{4 (FISTA $\lambda = 10^{-2}$)} & 100000 & 105891.07 & 5.89\% \\
            \added{4 (NAG)} & 100000 & 104144.20 & 4.14\% \\
            \hline
            \hline
        \end{tabular}}
\end{table}

\added{The high relative error and partially incorrect source localization seen in source 4 using FISTA is due to the current values set for the parameters in the minimization. The presence of the $l_{1}$ term attempts to induce sparsity in the solution of Eq.~\ref{eq:mmse}~\cite{fista2009}. Thus, as the minimization undergoes more iterations, the optimal solution will contain lower source activity in certain pixels. Adjusting the regularization parameter $\lambda$ and the total number of iterations before terminating the minimization will improve the calculated source activity and location with FISTA. However, we kept the FISTA reconstruction parameters consistent in this analysis and optimized them based on source 2.}

\begin{figure}[!htbp]
    \captionsetup{font=footnotesize}
        \centering
        \includegraphics[width=1\linewidth]{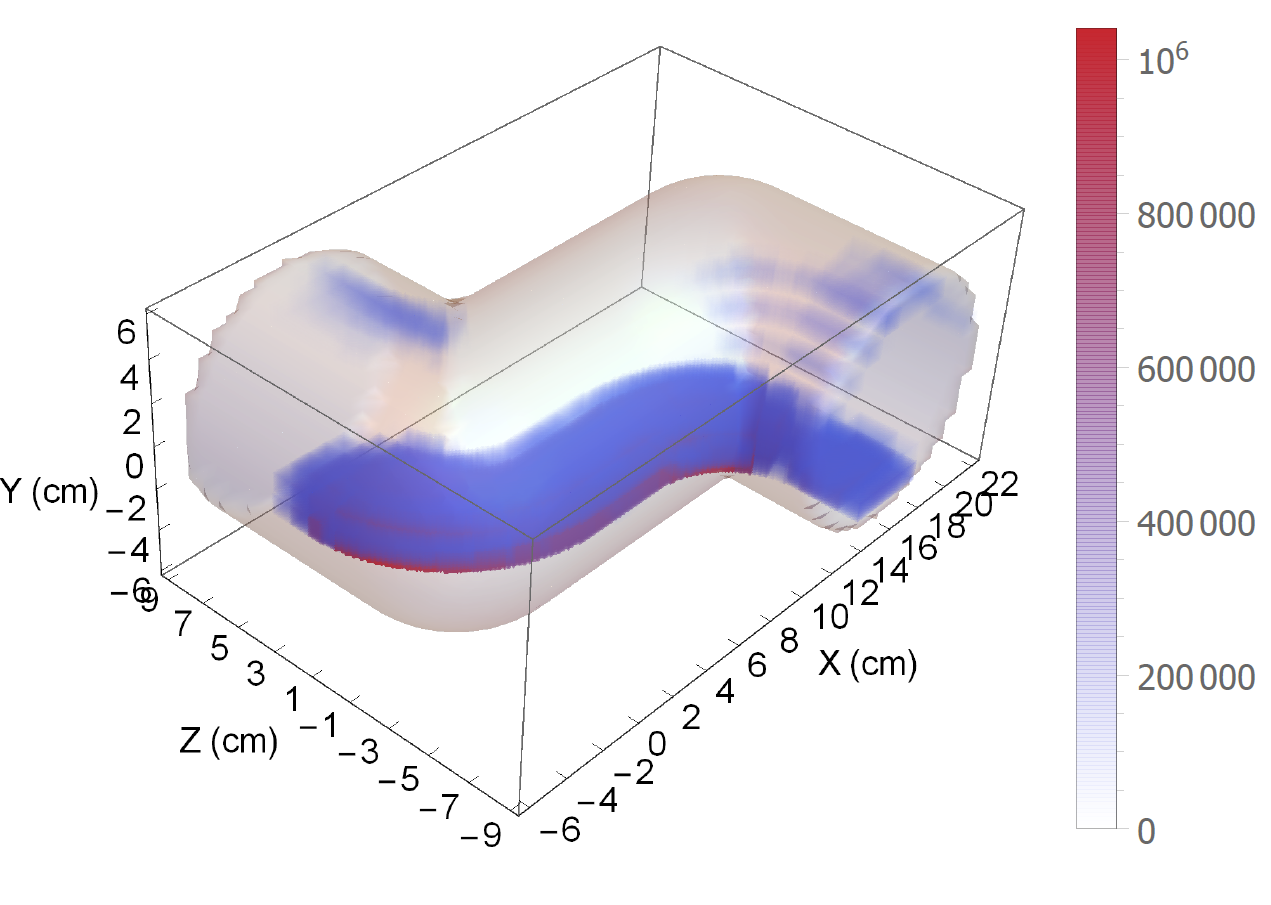}
        \caption{\added{Source 4 (FISTA with $\lambda = 10^{-2}$ and 30 iterations)}}
        \label{fig:3d_resultsource4FISTA}
    \end{figure}

\added{To improve on the results seen in source 4 using FISTA, we tested different reconstruction parameters and found that using $\lambda = 10^{-2}$ and stopping minimization at 30 iterations would result in a favorable source distribution well resembling the simulated one as seen in Fig.~\ref{fig:3d_resultsource4FISTA}, with the calculated activity at Table~\ref{table:activity estimate 3D}.}

\added{In addition, we believed it would be beneficial to show that using a different method can possibly improve reconstruction results of specific sources. An alternative gradient descent optimization algorithm, i.e., Nesterov accelerated gradient (NAG) descent, was used. Further details on NAG can be found at reference~\cite{NAG}.}
 
\begin{figure}[!htbp]
    \captionsetup{font=footnotesize}
        \centering
        \includegraphics[width=1\linewidth]{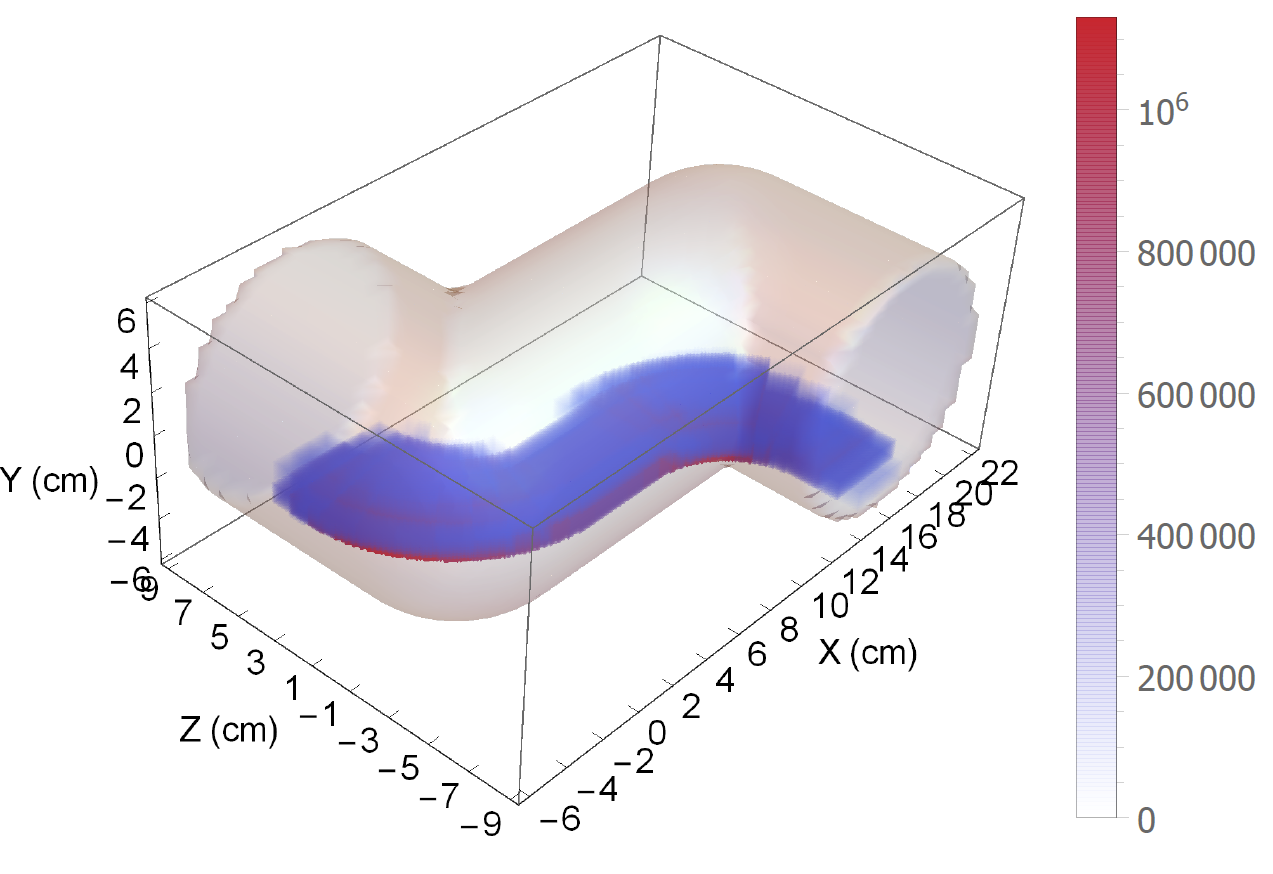}
        \caption{\added{Source 4 (NAG)}}
        \label{fig:3d_resultsource4NAG}
    \end{figure}

\added{The use of the NAG algorithm resulted in a smaller relative error in the calculated source activity and a cleaner, more accurate source localization as shown in Table~\ref{table:activity estimate 3D} and Fig.~\ref{fig:3d_resultsource4NAG}, respectively.}

\newpage

\section{Conclusions}
In this work, we demonstrated that current methods of holdup assay based on the GGH model could lead to significant underestimation of holdup mass due to the oversimplification of source geometry. We have developed an inverse-problem based approach, utilizing an array of detectors that rely on the accurate modeling of a system response matrix to reconstruct an image of the source distribution and estimate the source activity to measure the holdup material. We were able to accurately determine the source location in the experiment, and both the source location and activity in the 2D simulation. Similarly, we extended the notion of the response matrix from 2D to 3D, obtaining an accurate estimation of both the source location and activity in varied 3D simulations. 
\added{The maximum activity discrepancy was reported in source 4, which is due to the current values set for the parameters in the minimization. Sources inside other pipe geometries could be simulated to test this occurrence with FISTA further, with a possible solution of setting different parameters in FISTA's minimization depending on the measured geometry of the pipe.} Conversely, the method was always able to localize the source accurately within the response matrix's spatial resolution. In principle, smaller voxels could be used to achieve a more accurate source localization. However, this would entail the solution of a linear system with a higher number of unknowns.  
\added{Therefore, an array encompassing more detectors may be needed to improve the spatial resolution. The detectors' size may need to be reduced to keep the current system form factor. Robotic automated motion of the detector array would be particularly advantageous in this case since its proximity to the source and an adaptive speed would allow to accumulate the needed integral counts needed to achieve a fine 3D reconstruction. Minimally tethered soft robots~\cite{softrobot2} equipped with such array seem particularly suitable to accomplish this goal, guaranteeing an optimal solid angle regardless of the size or shape of the item inspected by crawling over the surface of the piece of equipment.}

Our method can be generalized to pipes of various curvatures and holdup of varying geometries, implementable into a minimally-tethered soft robot capable of reducing the manual labor and cost associated with holdup measurement. Using an i7-8565U processor, the process of minimization using FISTA for all tested sources had a maximum run-time of approximately \added{4} seconds, \added{heavily dependant on the number of measurements taken before minimization}. Given the fast convergence, solving the source distribution and activity requires little computational cost and could be implemented on off-the-shelf electronic platforms enabling real-time holdup imaging inside a fuel processing facility.


\section*{Acknowledgements}
This work is funded in-part by the Consortium for Verification Technology under Department of Energy National Nuclear Security Administration award number DE-NA0002534 and by the Nuclear Regulatory Commission Faculty Development Grant number 31310019M0011.

\section*{References}

\bibliography{references}

\end{document}